\documentclass[letterpaper]{article}
\usepackage{isea}
\usepackage[pdftex]{graphicx}
\usepackage{times}
\usepackage{helvet}
\usepackage{courier}
\usepackage[numbers]{natbib}
\usepackage{url}
\usepackage{balance}
\title{Robotic Blended Sonification: Consequential Robot Sound as Creative Material for Human-Robot Interaction}
\author{Stine S. Johansen, Yanto Browning, Anthony Brumpton, Jared Donovan, Markus Rittenbruch \\
Queensland University of Technology\\
Brisbane, Australia\\
\{stine.johansen, y.browning, a.brumpton, j.donovan, m.rittenbruch\}@qut.edu.au\\
\newline
\newline
}
\begin{document} 
\maketitle
\begin{abstract}

Current research in robotic sounds generally focuses on either masking the consequential sound produced by the robot or on sonifying data about the robot to create a synthetic robot sound. We propose to capture, modify, and utilise rather than mask the sounds that robots are already producing. In short, this approach relies on capturing a robot’s sounds, processing them according to contextual information (\textit{e.g.}, collaborators' proximity or particular work sequences), and playing back the modified sound. Previous research indicates the usefulness of non-semantic, and even mechanical, sounds as a communication tool for conveying robotic affect and function. Adding to this, this paper presents a novel approach which makes two key contributions: (1) a technique for real-time capture and processing of consequential robot sounds, and (2) an approach to explore these sounds through direct human-robot interaction. Drawing on methodologies from design, human-robot interaction, and creative practice, the resulting `Robotic Blended Sonification' is a concept which transforms the consequential robot sounds into a creative material that can be explored artistically and within application-based studies.

\end{abstract}


Robotics, Sound, Sonification, Human-Robot Collaboration, Participatory Art, Transdisciplinary

\section{Introduction and Background}

The use of sound as a communication technique for robots is an emerging topic of interest in the field of Human-Robot Interaction (HRI). Termed the ``Robot Soundscape'', Robinson et al. mapped various contexts in which sound can play a role in HRI. This includes ``\textit{sound uttered by robots, sound and music performed by robots, sound as background to HRI scenarios, sound associated with robot movement, and sound responsive to human actions}''~\cite[p. 37]{robinson2023robot}. As such, robot sound encompasses both semantic and non-semantic communication as well as the sounds that robots inherently produce through their mechanical configurations. With reference to product design research, the latter is often referred to as ``consequential sound''~\cite{van2008experience}. This short paper investigates the research question: How can consequential robot sound be used as a material for creative exploration of sound in HRI?

This research offers two key contributions: (1) an approach to using, rather than masking~\cite{trovato2018sound}, sounds directly produced by the robot in real-time, and (2) offering a way to explore those sounds through direct interactions with a robot. As an initial implication, this enables explorations of the sound through creative and open-ended prototyping. In the longer-term, this has the potential of leveraging and extending collaborators’ existing tacit knowledge about the sounds that mechanical systems make during particular task sequences as well as during normal operation versus breakdowns. Examples of using other communication modalities exist, mostly relying on visual feedback. Visual feedback allows collaborators to see, \textit{e.g.}, intended robotic trajectory and whether it is safe to move closer to the robot at any time. This assumes, however, that the human-robot collaboration follows a schedule in which the collaborator is aware of approximately when they can approach the robot. Sometimes, this timing is not possible to schedule, and collaborators must maintain visual focus on their task. This means that it is crucial to investigate ways of providing information about the robot’s task flow and appropriate timings for collaborative tasks. In other words, there is a need for non-visual feedback modalities that enable collaborators to switch between coexistence and collaboration with the robot. In order to achieve this aim, it is necessary to make these non-visual modalities of robot interaction available for exploration as creative `materials' for prototyping new forms of human-robot interaction.

\begin{figure*}[t]
\includegraphics[width=\textwidth]{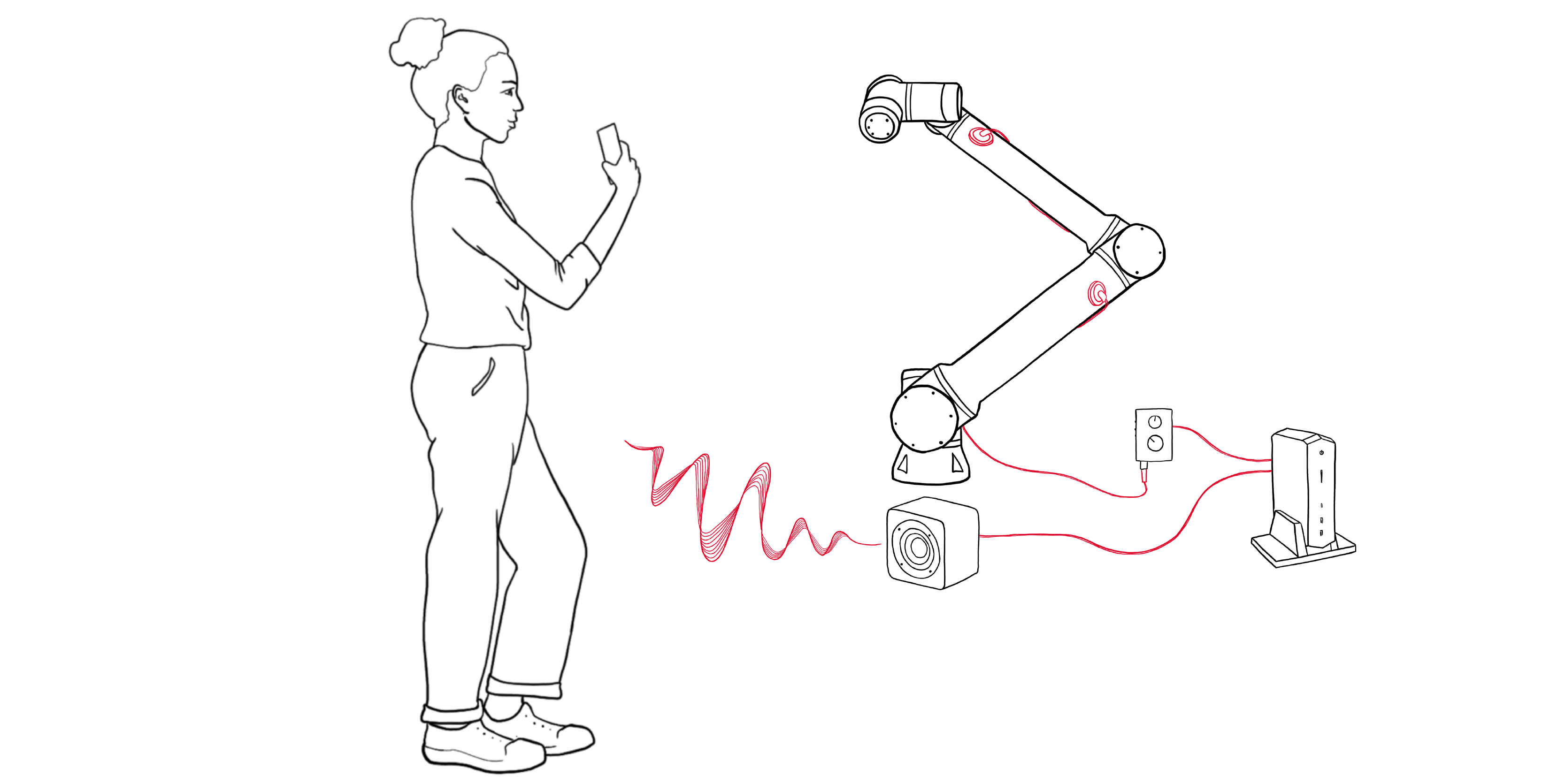}
\caption{The approach to Robotic Blended Sonification relies on (1) using electromagnetic field microphones that capture audio vibrations through solid objects, (2) processing the audio, and (3) reproducing the audio in real-time as the collaborator interacts with the robot by guiding the position of the TCP.}
\end{figure*}

Prototyping sound design for social robots has received particular attention in prior research, \textit{e.g.}, movement sonification for social HRI~\cite{frid2018perception}. However, this knowledge cannot be directly transferred when designing affective communication, including sound, for robots that are not anthropomorphic, \textit{e.g.}, mobile field robots, industrial robots for manufacturing, and other typical utilitarian robots~\cite{bethel2006auditory}. In prior research of consequential robot sound, Moore et al. studied the sounds of robot servos and outlined a roadmap for research into ``\textit{consequential sonic interaction design}''~\cite{Moore2017}. The authors state that robot sound experiences are subjective and call for approaches that address this rather than, \textit{e.g.}, upgrade the quality of a servo to reduce noise objectively. Frid et al. also explored mechanical sounds of the Nao robot for movement sonification in social HRI~\cite{frid2018perception}. They evaluated this through Amazon Mechanical Turk, where participants rated the sounds according to different perceptual measures Extending this into ways of modifying robot sounds, robotic sonification that conveys intent without requiring visual focus has been created by mapping movements in each degree of freedom for a robot arm to pitch and timbre~\cite{zahray2020robot}. The sound in that study, however, was created from sample motor sounds as opposed to the actual and real time consequential sounds of the robot. Another way this has been investigated is with video of a moving robot, Fetch, overlaid with either mechanical, harmonic, and musical sound to communicate the robot's inner workings and movement~\cite{Robinson2021}. This previous research indicates that people can identify nuances of robotic sounds but has yet to address if that is also the case for real time consequential robot sounds.

\section{Robotic Blended Sonification}

Robot sound has received increasing interest throughout the past decade, particularly for designing sounds uttered or performed by robots, background sound, sonification, or masking consequential robot sound~\cite{trovato2018sound}. Extending this previous research, we contribute with a novel approach to utilising and designing with consequential robot sound. Our approach for `Robotic Blended Sonification' bridges prior research on consequential sound, movement sonification, and sound that is responsive to human actions. Furthermore, it relies on the real-time sounds of the robot as opposed to pre-made recordings that are subsequently aligned to movements. A challenge for selecting the sounds a robot could make is that people have a strong set of pre-existing associations between robots and certain kinds of sounds. On one hand, this might provide a basis for helping people to interpret an intended meaning or signal from a sound (\textit{e.g.}, a danger signal), but it also risks that robot sounds remain cliched (beeps and boops), and may ultimately limit the creative potentials for robotic sound design. In this sense, Robotic Blended Sonification is an appealing approach because it offers the possibility of developing a sonic palette grounded in the physical reality of the robot, while also allowing for aspects of these sounds to be amplified, attenuated, or manipulated to create new meanings. Blended sonification has previously been described as ``\textit{the process of manipulating physical interaction sounds or environmental sounds in such a way that the resulting sound signal carries additional information of interest while the formed auditory gestalt is still perceived as coherent auditory event}''~\cite{tunnermann2013blended}. As such, it is an approach to augment existing sounds for purposes such as conveying information to people indirectly.

To achieve real-time robotic blended sonification, we use a series of electromagnetic field microphones placed at key articulation points on the robot. Our current setup uses a Universal Robots UR10 collaborative robotic arm. The recorded signals are amplified and sent to a Digital Audio Workstation (DAW), where they can be blended with sampled and synthesized elements and processed in distinct ways to create interactive soundscapes. Simultaneously to the real-time capture of the robot's audio signals, we enable direct interactions with the robot through the Grasshopper programming environment within Rhinoceros 3D (Rhino) and the RobotExMachina bridge and Grasshopper plugin~\cite{castillomachina}. We capture the real-time pose of the robot's Tool Center Point (TCP) in Grasshopper. Interaction is made possible via the Open Sound Control (OSC) protocol, with the Grasshopper programming environment sending a series of OSC values for the TCP. The real-time positional data also includes the pitch, roll, and yaw of each section of the robotic arm. Interaction with the robot arm is enabled through the Fologram plugin for Grasshopper and Rhino. The virtual robot is anchored to the position of the physical robot. The distance between the base of the robot and a smartphone is then calculated and used to direct the TCP towards the collaborator. This enables real-time interaction for exploring sounds for different motions and speeds.

For our prototype, OSC messages from the robotic movements are received in the Ableton Live DAW, along with the Max/MSP programming environment, and then assigned to distinct parameters of digital signal processing tools to alter elements of the soundscape. The plan for the initial prototype setup is to use five discrete speakers: A quadraphonic setup to allow for 360 degree coverage in a small installation space, along with a point source speaker located at the base of the robotic arm. The number of speakers is scalable to the size of the installation space and intent of the installation. The point source speaker alone is enough to gather data on the effects of robotic blended sonification on HRI, while multi-speaker configurations allow for better coverage in larger environments, enable investigations for non-dyadic human-robot interactions, and provide more creative options when it comes to designing soundscapes. 

\section{Directions for Future Research}

Ways of using non-musical instruments for musical expressions have a long history within sound and music art. Early examples include the work of John Cage, \textit{e.g.}, Child of Tree (1975) where a solo percussionist performs with electrically amplified plant materials~\cite{childoftree}, or the more recent concert Inner Out (2015) by Nicola Giannini where melting ice blocks are turned into percussive elements~\cite{innerout}. In a similar manner, our approach enables performance with robotic sound, subsequently allowing for a creative exploration of how those sounds affect and could be utilised for better human-robot collaborations. With the proposed approach, we identify new immediate avenues for research in the form of the following research questions:

\subsection{Robot Sound as Creative Material}

\textit{In what ways can the consequential sound of a robot be used as a creative material in explorations of robot sound design?} This can entail investigations through different configurations, including dyadic and non-dyadic interactions, levels of human-robot proximity, and different spatial arrangements. Furthermore, the interaction itself will play a crucial part in the way that the sound is both created and experienced, \textit{e.g.}, whether a collaborator is touching the robot physically or, as in our current setup, is interacting on a distance.

\subsection{Processing Consequential Robot Sound}

\textit{In what ways can or should we process the consequential sound material?} Two key points are connected to this. First, the consequential sound forms a basis for the resulting sound output which can be modified in various ways. Future research can entail exploring these, including the fact that different robots produce different consequential sounds that, subsequently, will lead to different meaningful modifications. Second, our approach can be complemented by capturing data from the surrounding environment to use as input for sound processing.

\subsection{Engaging People in Reflection}

\textit{How can we prompt people's reflections about consequential robot sounds through direct interaction?} While prior research has demonstrated ways to investigate consequential robot sound, \textit{e.g.}, through overlaying video with mechanical sounds, our approach enables people to explore sounds that result from their own interactions with a robot. This can be utilised for both structured and unstructured setups, depending on the purpose of the investigation. In our current setup, we invite for artistic exploration and expression. For more utilitarian purposes, the setup can be created in the context within which a robot is or could be present. This could support other existing methods for mapping and designing interventions into soundscapes.

\section{Conclusion}
In this short paper, we have described a novel approach for exploring and prototyping with consequential robot sound. This approach extends prior research by providing a technique for capturing, processing, and reproducing sounds in real-time during collaborators' interactions with the robot.

\section{Acknowledgments}
This research is jointly funded through the Australian Research Council Industrial Transformation Training Centre (ITTC) for Collaborative Robotics in Advanced Manufacturing under grant IC200100001 and the QUT Centre for Robotics.

\bibliographystyle{isea}
\bibliography{isea}

\balance

\section{Author Biographies}

\textbf{Stine S. Johansen} is a Postdoctoral Research Fellow in the Australian Cobotics Centre. Her research focuses on designing interactions with and visualisations of complex cyber-physical systems.

\vspace{0.2cm}

\noindent \textbf{Yanto Browning} is Lecturer at Queensland University of Technology in music and interactive technologies, with extensive experience as audio engineer.

\vspace{0.2cm}

\noindent \textbf{Anthony Brumpton} is artist academic working in the field of Aural Scenography. He likes the sounds of birds more than planes, but thinks there is a place for both.

\vspace{0.2cm}

\noindent \textbf{Jared Donovan} is Associate Professor at Queensland University of Technology. His research focuses on finding better ways for people to be able to interact with new interactive technologies in their work, currently focusing on the design of robotics to improve manufacturing.

\vspace{0.2cm}

\noindent \textbf{Markus Rittenbruch}, Professor of Interaction Design at Queensland University of Technology, specialises in the participatory design of collaborative technologies. His research also explores designerly approaches to study how collaborative robots can better support people in work settings.

\end{document}